\documentclass[twocolumn,showpacs,preprintnumbers,amsmath,amssymb,pra]{revtex4}
\usepackage{revsymb}
\usepackage{amstext,dcolumn,color}
\usepackage[pdftex]{graphicx}

\newcommand{\beq}{\begin{eqnarray}}
\newcommand{\eeq}{\end{eqnarray}}

\newcommand{\dev}{\mathrm d}
\newcommand{\I}{\imath}
\newcommand{\E}{\mathrm{e}^}

\newcommand{\re}{\mathrm{Re}}

\newcommand{\mS}{\mathcal{S}}
\newcommand{\mR}{\mathcal{R}}

\newcommand{\hc}{\mathrm{h.c.}}

\newcommand{\Q}{\dot{\mathcal{Q}}}

\begin{document}

\title{Spontaneous emission of an atom near a wedge}
\author{F.S.S. Rosa}
\affiliation{Theoretical Division, Los Alamos National Laboratory, Los Alamos, NM 87545, USA}
\author{T.N.C. Mendes}
\affiliation{Universidade Federal Fluminense - Niter\'oi, RJ, 24210-340,  Brazil}
\author{A. Ten\'orio}
\affiliation{Centro Brasileiro de Pesquisas F\'\i sicas - Rio de Janeiro, RJ, 22290-180, Brazil}
\author{C. Farina}
\affiliation{Universidade Federal do Rio de Janeiro - Rio de Janeiro, RJ, 21941-972 , Brazil}

\begin{abstract}
It is a well known fact that the presence of material
bodies in the vicinity of an atom affects its interaction with
the always present quantized electromagnetic field. In
this paper, we focus on how the spontaneous emission rate of a given
excited atom is altered when this atom is placed
inside a perfectly conducting wedge. We begin by briefly presenting
the formalism on which our calculations are founded, proceeding then
to a long but straightforward calculation of the transition rate. We
present results for a general atom but, for the sake of simplicity,
we narrow them down to an effective two-level system in our
numerical investigations. The oscillatory pattern for
 the spontaneous emission rate of the atom as we vary its relative
position to the wedge as well as the phenomenon of suppression are
shown in a couple of graphs.
\end{abstract}


\maketitle

\section{Introduction}

Last year we have celebrated the ninetieth anniversary of A.
Einstein's historical paper on his theory of
radiation\cite{Einstein}, the final paper of a series that began
more than a decade earlier with his 1905 work on the quantization of
the radiation field. By trying to explain how atoms could be in
thermal equilibrium with the radiation field, he introduced the
novel concepts of spontaneous and stimulated emission of radiation
\cite{footnote1}, that again outlined how unusual quantum mechanics
could be and reheated a dormant discussion about causality in the
quantum scale \cite{Kleppner}. However, despite its huge
achievements this remarkable work had at least one shortcoming,
since it did not furnish the means for evaluation of the spontaneous
emission rate. This had to wait another ten years, until P.A.M.
Dirac successfully calculated it in a paper that most regard as the
birth of quantum electrodynamics \cite{Dirac}.

Once established the framework of the modern quantum field theory, a
meaningful question to ask is how spontaneous emission (SE) was
affected by the presence of boundaries. That nontrivial boundary
conditions {\it could} modify the SE rate is very clear from the QED
standpoint, since it is the quantized electromagnetic field that
causes SE and the field is definitely affected by boundary
conditions \cite{footnote2}. This issue was firstly considered by
E.M. Purcell \cite{Purcell} in the forties, and subsequent
experimental work showed that SE could be strongly modified if
boundaries were present \cite{Feher, Drexhage}. The first explicit
theoretical results \cite{Barton, Stehle} for parallel conducing
plates also supported the conclusion that boundary conditions could
significantly enhance or suppress the SE rate
\cite{Hinds-LesHouches}.

In this work we carry on the analysis on how boundaries may
influence the SE process. We investigate the behavior of an atom
inside a wedge made of conducting plates. The reason why we chose
such a setup is twofold. Firstly, the wedge configuration is a
relatively simple system that allows the study of nontrivial
curvature effects in the zero-point field \cite{Saharian,
Deustch-Candelas, Brevik1, Brevik2}, due to the presence of the
cusp. In addition, the wedge geometry has already been used in
experiments \cite{Hinds}-although thereabouts the wedge aspect was
not important, it may probably be in the future. Such configuration
may also be relevant for trapping atoms in excited states, since a
suited geometry could lead to locations where the potential well
disturbing the atom is deep and the SE rate is small.

The plan of this paper is arranged as follows. In the next section
we present the necessary formalism in some detail, and show how it
leads to a systematic way of evaluating SE rates. In section III we
define our problem and proceed with the calculations needed in order
to bring the result to the most convenient form for numerical
investigations, leaving the fourth section to conclusions and
perspectives.


\section{Transition rates of a system coupled to a reservoir}

\subsection{General formalism}

Let be a small system $\mathcal S$ characterized by a time dependent
density matrix operator $\rho_S(t)$ which interacts with a large
system $\mathcal R$ which can be treated as a reservoir and
characterized by a time independent density matrix operator
$\rho_R$. The Hamiltonian of the whole system $\mathcal S+\mathcal
R$ is $H=H_S+H_R+V$, where $H_S$($H_R$) is the unperturbed
Hamiltonian of $\mathcal S$($\mathcal R$) with eigenstates $\vert
a\rangle$($\vert\mu\rangle$) and with eigenvalues $E_a$($E_{\mu}$).
By hypothesis, the interaction hamiltonian between $\mathcal S$ and
$\mathcal R$ is
\beq
\label{V}
V=\sum_jS_jR_j\,,
\eeq
where $S_j$ and $R_j$ are compatible observables associated to
$\mathcal S$ and $\mathcal R$ respectively. Assuming that
correlations occurring between $\mathcal S$ and $\mathcal R$ last a
time $\tau_c$ and that variations of the observables in $\mathcal S$
are characterized by a time scale $T_S$, one is able to write a {\it
coarse grained rate} equation for the time evolution of the diagonal
elements of $\rho_S(t)$ in the base ket $\vert\{a\}\rangle$
\cite{Cohen1982,Cohen1984},
\beq
\label{EQMP}
{\dev\over \dev t}\rho_{aa}^S\left(t\right)
&=&
 \sum_{c}\left(\rho_{cc}^S\left(t\right)\Gamma_{c\rightarrow a} -
 \rho_{aa}^S\left(t\right)\Gamma_{a\rightarrow c}\right) \, ,
\eeq
where $dt$ is not infinitesimal but satisfies the inequality
$\tau_c\ll dt \ll T_S$, provided of course that such a relation
holds for some $dt$. The factor $\Gamma_{c\rightarrow a}$ is defined
by
\beq
\label{Fermi}
 \Gamma_{c\rightarrow a} \hspace{-3pt}&=&
 \hspace{-3pt} {2\pi\over\hbar}\sum_{\mu}p_{\mu}\sum_{\nu}\vert
 \langle \mu,c\vert V\vert\nu,a\rangle\vert^2  \cdot \nonumber \\
 &&\;\;\;\;\;\;\;\;\;\;\; \delta\left(E_{\mu}+E_c-E_{\nu}-E_{a}\right) \, ,
\eeq
where $\rho_{ab}^S = \langle a\vert\rho_S\vert b\rangle$ and
$p_{\mu}$ is the statistical weight for a state $\vert\mu\rangle$ of
the reservoir. We see that $\Gamma_{c\rightarrow a}$ may be
interpreted as the probability (per unit time) of a transition in
the system between the states $\vert c\rangle$ and $\vert a\rangle$.
Equation (\ref{Fermi}) is precisely the Fermi's {\it golden rule}.

In order to obtain the exchange energy rates, one must consider the
time derivative of the average value of the hamiltonian of the
system. For a given eigenstate $\vert a \rangle$ of the system, a
simple application of Eq. (\ref{EQMP}) shows that the net exchange
energy rate between $\mS$ and $\mR$ is
\beq
\label{txHa}
{\dev\langle H_S\rangle_a\over \dev t} = \dot\rho_{aa}^SE_a =
 -\sum_{b}\hbar\omega_{ab}\Gamma_{a\rightarrow b} \, ,
\eeq
where $\omega_{ab} = (E_a - E_b)/\hbar$. Following the work
developed by J. Dalibard {\it et al.}  \cite{Cohen1982,Cohen1984}, we now proceed to recast
the previous result into a more intuitive form that clearly exhibits
the roles played by $\mS$ and $\mR$. Let us begin by defining the
symmetric correlation function $C_{jk}^R\left(\omega\right)$ and the
linear susceptibility $\chi_{jk}^R\left(\omega\right)$ of the
reservoir in the frequency space \cite{AtPhInt,Kubo66},
{\small \beq
\label{FC}
\hat C_{jk}^R\left(\omega\right) &=& \pi \sum_{\mu}p_{\mu} \cdot \nonumber \\
&& \! \sum_{\nu}R_{\mu\nu}^{j} R_{\nu\mu}^{k}
 \Big[ \delta\left( \omega+\omega_{\mu\nu}\right) +\delta\left( \omega-\omega_{\mu\nu}\right) \Big] \, ,
\\
%
%
\hat\chi_{jk}^R\left(\omega\right) &=&
\hat\chi_{jk}^{\,\prime R}\left(\omega\right) +
 \I\,\hat\chi_{jk}^{\,\prime\prime R}\left(\omega\right) \, ,
\\
%
\label{Chi'R}
\hat\chi_{jk}^{\,\prime R}\left(\omega\right) &=&
 -\frac{1}{\hbar}\sum_{\mu}p_{\mu}  \cdot \nonumber \\
&& \! \sum_{\nu}R_{\mu\nu}^{j} R_{\nu\mu}^{k}
  \left[ {\mathcal P}\frac{1}{\omega_{\mu\nu}+\omega}+{\mathcal P}\frac{1}{\omega_{\mu\nu}-\omega}\right] \, ,
\\
%
\label{Chi''R}
\hat\chi_{jk}^{\,\prime\prime R}\left(\omega\right) &=&
 \frac{\pi}{\hbar}\sum_{\mu}p_{\mu}  \cdot \nonumber \\
&& \! \sum_{\nu}R_{\mu\nu}^{j} R_{\nu\mu}^{k}
 \Big[ \delta\left(\omega_{\mu\nu}+\omega\right)-\delta\left(\omega_{\mu\nu}-\omega\right)\Big] \, ,
\eeq}
where $R_{\mu\nu}^j=\langle\mu\vert R_j\vert\nu\rangle$, $\omega_{\mu\nu}=\left(E_{\mu}-E_{\nu}\right)/\hbar$ and $\mathcal P$ is the principal part. The equivalent expressions for the system $\mathcal S$ are
{\small \beq
\label{CSw}
\!\!\!\!\!\hat C_{jk}^{S,a}\left(\omega\right) &=&
 \pi \sum_{n} S_{a n}^{j} S_{n a}^{k}\Big[ \delta\left( \omega+\omega_{an}\right) +
 \delta\left( \omega-\omega_{an}\right) \Big] \;\;\;\;
\\
%
\label{Chi'S}
\!\!\!\!\!\hat \chi_{jk}^{\,\prime S,a}\left(\omega\right) &=&
 -\frac{1}{\hbar}\sum_{n} S_{a n}^{j} S_{n a}^{k}\left[ {\mathcal
P}\frac{1}{\omega_{a n}+\omega}+{\mathcal P}\frac{1}{\omega_{a
n}-\omega}\right] \;\;\;\;
\\
%
\label{Chi''S}
\hat\chi_{jk}^{\,\prime\prime\, S,a}\left(\omega\right) &=&
 \frac{\pi}{\hbar}\sum_{n}S_{an}^{j} S_{na}^{k}\Big[ \delta\left(\omega_{an} +
  \omega\right)-\delta\left(\omega_{an}-\omega\right)\Big] \;\;\;\;
\eeq}
with $S_{ab}^j=\langle a\vert S_j\vert b\rangle$. It is now a matter of straightforward algebra to show that (\ref{txHa}) may be rewritten in terms of these new quantities
\beq
\label{Qa}
{\dev\langle H_S\rangle_a\over \dev t} &=& \Q_a\; =\; \dot{\mathcal Q}_a^{fr} + \dot{\mathcal Q}_a^{rr} \, ,
\\
%
\label{dfr}
\dot{\mathcal Q}_a^{fr} &=&
\sum_{j,k}\int_{-\infty}^{\infty}{\dev\omega\over 2\pi}\,\omega\,\hat\chi_{jk}^{\,\prime\prime\, S,a}
 \left(\omega\right)\hat C_{kj}^{R}\left(\omega\right) \, ,
\\
%
\label{drr}
\dot{\mathcal Q}_a^{rr} &=&
 -\sum_{j,k}\int_{-\infty}^{\infty}{\dev\omega\over 2\pi}\,\omega\,\hat\chi_{jk}^{\,\prime\prime\,R}
  \left(\omega\right)\hat C_{kj}^{S,a}\left(\omega\right) \, .
\eeq
The physical interpretations of $\dot{\mathcal Q}_a^{fr} $ and
$\dot{\mathcal Q}_a^{rr}$ are simple. The former may be understood
as the power kicked into the system $\mS$ by the fluctuations of the
reservoir $(fr)$, while the latter represents the power lost to the reservoir due to fluctuations of the system itself. This last term is also called the reservoir reaction contribution
$(rr)$ since these fluctuations of the system affect the reservoir, which then back reacts on the system \cite{Cohen1982,Cohen1984}.

\subsection{Dipole interacting with the radiation field}

Let us now take our system to be a small neutral entity,
characterized by the set of transition frequencies $\omega_{ba}$,
and let us assume that it is in contact with the quantized radiation
field, taken as the reservoir. The radiation eletric field may be
written as
\beq
\label{Efield}
\mathbf{E}\left(\mathbf{x},t\right) =
 \sum_{\mathbf{k}\lambda}\left(\mathbf{f}_{\mathbf{k}\lambda}
 \left(\mathbf{x}\right)\E{\I\omega_k t} a_{\mathbf{k}\lambda}^{\dag}+\mathbf{f}_{\mathbf{k}\lambda}^{*}
 \left(\mathbf{x}\right)\E{-\I\omega_k t} a_{\mathbf{k}\lambda}\right)
\eeq
where each mode is specified by a wavevector $\mathbf{k}$ and
polarization $\lambda$, with  $\omega_k=c\vert\mathbf{k}\vert$.  The
annihilation and creation operators $a_{\mathbf{k}\lambda}$ and
$a_{\mathbf{k}\lambda}^{\dag}$ satisfy the usual commutation
relations
\beq
\label{Com_aa}
\big[a_{\mathbf{k}\lambda},a_{\mathbf{k}^{\prime}\lambda^{\prime}}\big] =
 \big[a_{\mathbf{k}\lambda}^{\dag},a_{\mathbf{k}^{\prime}\lambda^{\prime}}^{\dag}\big] = 0 \, ,\\
\label{Com_aa_2}
\big[a_{\mathbf{k}\lambda},a_{\mathbf{k}^{\prime}\lambda^{\prime}}^{\dag}\big] =
 \delta_{\mathbf{k}\mathbf{k}^{\prime}}\delta_{\lambda\lambda^{\prime}} \, ,
\eeq
and the function $\mathbf{f}_{\mathbf{k}\lambda}\left(\mathbf{x}\right)$ is the position dependent part of the mode $\mathbf{k}\lambda$ and carry the information about the boundary conditions and possible source
contributions.

Let us assume that the dimensions of the system $\mathcal S$
are too small compared to any of its transition
wavelengths, that is, $a_0\ll 2\pi c/\omega_{ab}$, where $a_0$ is
the largest dimension of the system, and also that $\mathcal S$ is
non-relativistic. Then the interaction part of the hamiltonian may
be approximated by the dipole interaction of a particle and the
electric field
\beq
\label{Vint_mode}
V\left(\mathbf{x},t\right) \hspace{-7pt}&&= -\mathbf{d}\cdot \mathbf{E}\left(\mathbf{x},t\right) = \nonumber \\
&&-\sum_{\mathbf{k}\lambda}\sum_{j} \left(d_{j} f_{\mathbf{k}\lambda}^{j}
\left(\mathbf{x}\right)\E{\I\omega_k t} a_{\mathbf{k}\lambda}^{\dag}+\hc\right) \, ,
\eeq
where $\mathbf{d}$ is the dipole moment operator of the system $f_{\mathbf{k}\lambda}^{j} \left(\mathbf{x}\right)$ is the $j$-component of $\mathbf{f}_{\mathbf{k}\lambda} \left(\mathbf{x}\right)$ and the index $j=1,2,3$ runs over the directional unitary vectors of a given tridimensional coordinate system. Assuming that the susceptibility and the correlation function of $\mathcal S$ are diagonal in the appropriate coordinate system, we have
\beq
\label{CSdjk}
&&\hat C_{jk}^{S,a}\left(\omega\right)\equiv \alpha^{(-)}_{aj}\left(\omega\right) \nonumber \\
&&=-\pi\hbar\sum_b{\alpha^{j}_{ab}\omega_{ab}\over 2}\Big[ \delta\left(\omega_{ab} +
 \omega\right)+\delta\left(\omega_{ab}-\omega\right)\Big] \, ,
\\
%
\label{XS''djk}
&&\hat\chi_{jk}^{\,\prime\prime S,a}\left(\omega\right)\equiv
 \alpha_{aj}^{(+)}\left(\omega\right) \nonumber \\
&&=\pi\sum_b{\alpha^{j}_{ab}\omega_{ab}\over 2}
 \Big[\delta\left(\omega_{ab}-\omega\right) -
\delta\left(\omega_{ab}+\omega\right)\Big] \, ,
\eeq
where
%
%
%
$ \alpha^{j}_{ab} =
-2\vert \langle a\vert
d_{j}\vert b\rangle\vert^2/\hbar\omega_{ab} $.
%

Using last equations in (\ref{dfr}) and (\ref{drr}), making
$\vert\mu\rangle\equiv\vert n_{\mathbf{k}\lambda}\rangle$ in
equation (\ref{Fermi}) (which means a Fock state with $n$ photons in
the mode $\mathbf{k}\lambda$) and performing the integration on
$\omega$, we obtain, after some calculations,
\beq
\label{Qa_rr}
\dot{\mathcal Q}_a^{rr} = - \sum_{j}\sum_{\mathbf{k}\lambda}c\,k\,\alpha_{aj}^{(-)}
 \left(k\right)\vert f^{j}_{\mathbf{k}\lambda}\left(\mathbf{x}\right)\vert^2 \, , \;\;\;\;
\\
%
\label{Qa_fr}
\dot{\mathcal Q}_a^{fr} =
\sum_{j} \sum_{\mathbf{k}\lambda} c\,k\,\alpha_{aj}^{(+)}\left(k\right)\vert f^{j}_{\mathbf{k}\lambda}
 \left(\mathbf{x}\right)\vert^2 \Big( 2\langle n_{\mathbf{k}\lambda}\rangle + 1\Big) \, ,
\eeq
%
where $\langle n_{\mathbf{k}\lambda}\rangle$ is the average number of photons in the mode $\mathbf{k}\lambda$.

As pointed out before, equation (\ref{Qa_rr}) gives the contribution
arising from the radiation reaction to the energy rates of the
system, while equation (\ref{Qa_fr}) gives the corresponding
contribution coming from the field fluctuations. From (\ref{Qa_rr})
and (\ref{CSdjk}), it is possible to see that the $(rr)$ contribution is never
positive, which means that it always accounts for power that is
emitted by the system. Since it also does not depend on $\langle
n_{\mathbf{k}\lambda}\rangle$, we conclude that the $(rr)$
contribution comes from {\it spontaneous} processes only. On the
other hand, Equations (\ref{Qa_fr}) and
(\ref{XS''djk}) show that the $(fr)$ contribution can have both
signs, meaning that $\dot{\mathcal Q}_a^{fr}$ could represent either
an emitted or an absorbed power. In addition, its dependence on
$\langle n_{\mathbf{k}\lambda}\rangle$ signals that {\it stimulated}
processes also play a role in the $fr$ contribution. For the special
case where $\langle n_{\mathbf{k}\lambda}\rangle=0$, there is no
absorbed power and the total (spontaneous) radiated power by the
system is
\beq
\label{Qnet}
\Q_a\hspace{-10pt}&&=\sum_{b,j}\Q_{ab,j} \nonumber \\
 \hspace{-10pt}&&= \pi c\sum_{j}\sum_{b<a}\alpha_{ab}^{j}k_{ab}\sum_{\mathbf{k}\lambda}k\vert f^{j}_{\mathbf{k}\lambda}
 \left(\mathbf{x}\right)\vert^2\delta\left( k - k_{ab}\right) \, , \;\;\;\;
\eeq
where $k_{ab} = \omega_{ab}/c$ and. Let us also note that the emitted power at a given permitted transition, $\Q_{ab,j}$, is proporcional to the spontaneous emission rate $\Gamma_{a\rightarrow b}^{j}$ characteristic of this transition.


\section{The spontaneous emission of an atom inside a wedge}

As far as we know, the method described in the previous section was firstly applied in atom-surface problems by D. Meschede {\it et al.} \cite{Meschede}, where they found the level shifts and the radiation rates of an atom near a single wall. A few years later, some of the present authors revived this method to evaluate thermal corrections to the same problem \cite{TarciroFarinaJPA2007}, and also to calculate the van der Waals interaction between an atom and a perfectly conducting wedge \cite{FLNS_JPA2008}. In this paper, we intend to continue this work by evaluating the spontaneous emission rates for the atom-wedge system. 

Let us consider the system described schematically in figure 1. It consists of an
atom placed inside of a perfectly conducting wedge,
characterized by a radius R and an \linebreak angle $\alpha$.  The location of the atom is given by its distance $\rho$ to the wedge axis and its declination $\varphi$ relative to the bisector plane, related to the auxiliary angle $\phi$ by $\varphi = \phi - \alpha/2$. Let us further assume that this atom is characterized by a dipole moment ${\bf d} = -e {\bf r}_e$, where $e$ is the fundamental charge and ${\bf r}_e$ is the electron position with respect to the center of the atom. It is then clear that taking the atom as our system $\mathcal S$ and the electromagnetic (EM) field constrained by the presence of the wedge as the reservoir $\mathcal R$,  we may straightforwardly apply the formalism discussed in the previous section.


\begin{figure}[h!]
\begin{center}
\scalebox{0.45}{\includegraphics{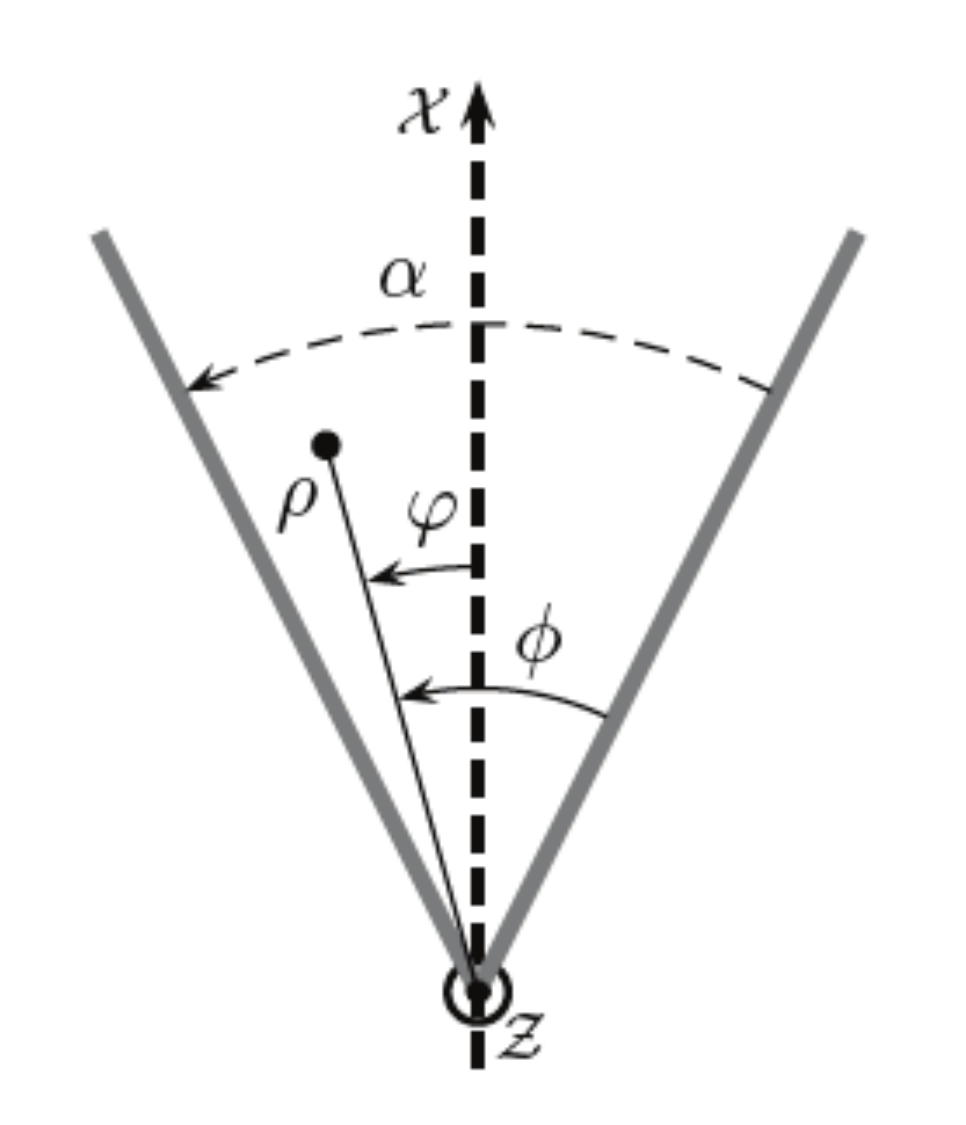}}
\caption{The Atom-wedge system. The atom is represented by the black circle.}
\end{center}
\label{Figure1}
\end{figure}

 %
%
%

According to (\ref{Qa_rr})-(\ref{Qa_fr}), we need the modes of the EM
field inside the wedge in order to evaluate the spontaneous emission
rate. As they have been already obtained, for instance, in
\cite{Saharian}, we will merely quote them
\begin{eqnarray}
\label{TM}
\mathbf{f}_{\mathbf{k},m,n}^{\mathrm{TM}}\left(\mathbf{x}\right)
\hspace{-3pt}&=&\hspace{-3pt}
\beta_{qm}\left(\gamma_{m,n}R\right)\left(\gamma_{
m,n}^2\hat{z} - \imath k_z\mathbf{\nabla}_t\right) \cdot \nonumber \\
&&
 \!\!\!\!\!\left[J_{q m}\left(\gamma_{m,n}\rho\right)\sin\left(q
m \phi\right)e^{-\imath k_z z}\right] \, ,
\\
%
\label{TE}
\mathbf{f}_{\mathbf{k},m,n}^{\mathrm{TE}}\left(\mathbf{x}\right) \hspace{-4pt}&=&\hspace{-4pt} \imath k\beta_{qm}\left(\eta_{ m,n}R\right) \cdot \nonumber \\
\hat{z}\times\mathbf{\nabla}_t&& \hspace{-20pt} \left[J_{q m}\left(\eta_{ m,n}\rho\right)\cos\left(q  m \phi\right)e^{-\imath k_z z}\right] \, ,
\end{eqnarray}
where  $q=\pi/\alpha$, $m \in \mathbb{N}^{*}$ in (\ref{TM}), $m \in \mathbb{N}$ in (\ref{TE}) and
\beq
\omega_k^2 =  c^2 \left[\gamma_{m ,n}^2 + k_z^2\right] \, .
\eeq
The quantities $\gamma_{m,n}$ and $\eta_{m,n}$ are defined by
\begin{eqnarray}
&&J_{q m}\left(\gamma_{ m,n}R\right) = J^{\,\prime}_{q m}\left(\eta_{m,n}R\right) = 0 \;\;\;\;\;\; n \in \mathbb{N}
\eeq
and
\beq
&&{\beta}_{qm}^2\left(x\right) = \frac{4q\hbar c}{\pi k} X_{qm}(x) \;\; \mbox{if} \;\; m \neq 0 \, ,\\
&&{\beta}_{qm}^2\left(x\right) = \frac{2q\hbar c}{\pi k} X_{qm}(x)
\;\; \mbox{if} \;\; m = 0 \, , \\
&&X_{\nu}(x) = \left[J^{\,\prime\,2}_{\nu}\left(x\right)+\left(1-\nu^2/x^2\right)J_{\nu}^2\left(x\right)\right]^{-1} \, , \\
&&\mathbf{\nabla}_t = \hat\rho\partial_{\rho}+\rho^{-1}\hat\phi\partial_{\phi} \, ,\\
&& k^2 = \kappa_{mn\lambda}^2+k_z^2 \, ,
\end{eqnarray}
with $\kappa_{mn0} = \gamma_{m,n}$  and $\kappa_{mn1} = \eta_{m,n}$. Inserting (\ref{TM})-(\ref{TE}) into (\ref{Qnet}) and setting $\langle n_{k \lambda} \rangle = 0$, since we are at zero temperature, we get after a little manipulation
\begin{widetext}
\begin{equation}
\label{EEa_1}
\Q_a =  2 \hbar \pi c^2 \sum_{\lambda,j} \sum_{b<a} \alpha_{ab}^{j}k_{ab} \int_{-\infty}^{\infty}\dev k_z \!\!\sum_{m=-\infty}^{\infty}\sum_{n=1}^{\infty}\kappa_{\vert m\vert n\lambda}^4 \, \delta \! \left(k_{ab} - \sqrt{\kappa_{\vert m\vert n\lambda}^2+k_z^2}\right)  X_{q\vert m\vert}\left(\kappa_{\vert m\vert n\lambda}R\right)Q_{q\vert m\vert,n}^{\,j ,\lambda}\left(\rho,\phi\right) \, , \;\;\;
\end{equation}
where $\lambda =0,1$, $j=r,\phi,z$, and
\begin{eqnarray}
\label{Q_pTM}
&&Q_{qm}^{\, \rho , 0}\left(\kappa_{mn0},\rho,\phi\right) = {k_z^2\over\kappa_{mn0}^2}J_{qm}^{\,\prime\,2}\left(\kappa_{mn0}\rho\right) \sin^2\left(qm\phi\right) \, ,\\
\label{Q_fiTM}
&&Q_{qm}^{\, \phi , 0}\left(\kappa_{mn0},\rho,\phi\right) = {k_z^2q^2 m^2\over \kappa_{mn0}^4\rho^2} J_{qm}^2\left(\kappa_{mn0}\rho\right) \cos^2\left(qm\phi\right) \, ,\\
\label{Q_zTM}
&&Q_{qm}^{\, z , 0}\left(\kappa_{mn0},\rho,\phi\right) =  J_{qm}^2\left(\kappa_{mn0}\rho\right)\sin^2\left(qm\phi\right) \, ,\\
\label{Q_pTE}
&&Q_{qm}^{\,\rho ,1} \left(\kappa_{mn1},\rho,\phi\right) = \left(1+{k_z^2\over\kappa_{mn1}^2}\right){q^2 m^2\over \kappa_{mn1}^2\rho^2} J_{qm}^2\left(\kappa_{mn1}\rho\right) sin^2\left(qm\phi\right) \, ,\\
\label{Q_fiTE}
&&Q_{qm}^{\, \phi , 1} \left(\kappa_{mn1},\rho,\phi\right) = \left(1+{k_z^2\over\kappa_{mn1}^2}\right)J_{qm}^{\,\prime\,2}\left(\kappa_{mn1}\rho\right) \cos^2\left(qm\phi\right) \, ,\\
\label{Q_zTE}
&&Q_{qm}^{\, z , 1}\left(\kappa_{mn1},\rho,\phi\right) = 0 \, .
\eeq

We may now use the generalized Abel-Plana formula
\cite{Saharian,Saharian2}, which is
\begin{eqnarray}
\sum_{n=1}^{\infty} (\kappa_{m n \lambda}R) X_{qm}(\kappa_{m n \lambda} R) f(\kappa_{m n \lambda}R)= \hspace{-5pt}&&\frac{1}{2} \int_0^{\infty} dx f(x) + \frac{\pi}{4} Res \left[ f(z) \frac{Y^{(\lambda)}_{qm}(z)}{J^{(\lambda)}_{qm}(z)}\right]_{z=0} \nonumber \\
\hspace{-5pt}&&-\frac{1}{2\pi} \int_0^{\infty} dx
\frac{K^{(\lambda)}_{qm}(x)}{I^{(\lambda)}_{qm}(x)} \left[ e^{-qm\pi
i}  f(e^{i \pi/2}x) + e^{qm\pi i} f(e^{-i \pi/2}x)\right] \, ,
\end{eqnarray}
where the superscript $(\lambda)$ in the Bessel functions mean their
$\lambda$-th derivative, in order to recast the expression \linebreak
(\ref{EEa_1}) into
\beq
\label{EEa_2}
\hspace{-0.5in}\Q_a &=& \hbar \pi c^2 \sum_{\lambda,j,b} \alpha_{ab}^{\,j}k_{ab} \int_{-\infty}^{\infty}\dev k_z \!\!\sum_{m=-\infty}^{\infty} \int_{0}^{\infty} \! dy \, y^3 \, \delta \! \left(k_{ab}-\sqrt{y^2+k_z^2}\right)  Q_{q\vert m\vert}^{j, \lambda}\left(y,\rho,\phi\right) \;+\; \mathcal{F}(R,\rho,\phi) \, ,
\eeq
\end{widetext}
where we were able to isolate all the R-dependence of $\Q_a$ into
an involved but given function $\mathcal{F}(R,\rho,\phi)$. As we are
interested in the wedge without the external cap, we must take the
$R \rightarrow \infty$ limit, and here is where the rearrangement we
made in (\ref{EEa_2}) shows its quality: it may be shown that
$\mathcal{F}(R \rightarrow \infty,\rho,\phi) \rightarrow 0$
\cite{Saharian}, leaving us only with the first term.

Although the generalized Abel-Plana formula made things simpler, expression (\ref{EEa_2}) still is not very efficient for numerical investigations. However, if we rewrite the opening angle as $\alpha = \pi/q$ and restrict ourselves to integer values of the parameter $q$, it turns out that further simplification is possible. In this case we can switch $\vert m \vert$ to $m$ in (\ref{EEa_2}) and carry out the $m$-summation by using an addition theorem relating Bessel functions \cite{Davies}
\begin{widetext}
\begin{equation}
\label{BesselGAT}
\sum_{m=-\infty}^{\infty}J_{qm}\left(\kappa\rho\right)Z_{\nu+qm}\left(\kappa\rho\right)\E{2\I qm\phi} =  {1\over q}\sum_{l=0}^{q-1}\left(-1\right)^{\nu/2}\E{-\I\nu\left(\phi + \frac{\pi l}{q}\right)}Z_{\nu}\left(2\kappa\rho\sin\left(\phi + \frac{\pi l}{q}\right)\right) \, ,
\end{equation}
\end{widetext}
where $Z_{\nu}$ is a given solution of Bessel equation. As slightly
different cases of this theorem will apply to each polarization, it
is convenient to split $\Q_a$ in its $\rho, \phi$ and $z$
contributions 
\beq 
\Q_a = \sum_{j} \Q_{a,j} = \Q_{a,\rho} + \Q_{a,\phi} + \Q_{a,z} 
\eeq
and consider each one separately. Beginning with the
$z$-contribution and substituting (\ref{Q_zTM}), (\ref{Q_zTE}) and
(\ref{BesselGAT}) into  $\Q_{a,z}$, we get
\begin{widetext}
\beq
\label{Q_a,z_1}
\hspace{-0.5in}\Q_{a,z} = \frac{\hbar \pi c^2}{2q} \sum_{b<a} \alpha_{ab}^{z}k_{ab} \int_{-\infty}^{\infty} \! &&dk_z  \int_{0}^{\infty} \! dy y^3 \, \delta \! \left(k_{ab}-\sqrt{y^2+k_z^2}\right) \cdot \nonumber \\
&&\sum_{l=0}^{q-1}\left[ J_0\left(2y\rho\sin\left(\frac{\pi l}{q}\right) \right) - J_0\left(2y\rho\sin\left(\phi + \frac{\pi l}{q}\right) \right) \right] \, .
\eeq
By performing the change of variables $y = k\sin\theta$, $k_z =
k\cos\theta$ and using the identity \cite{Gradshteyn}
\beq
\label{identityGradshteyn}
\int_0^{\pi/2}J_{\mu}\left(a\sin\theta\right)\left(\sin\theta\right)^{\mu+1}\left(\cos\theta\right)^{2\nu+1}\dev\theta = 2^{\nu}\Gamma\left(\nu+1\right)a^{-\nu-1}J_{\nu+\mu+1}\left(a\right) \, ,
\eeq
valid when $\re\left[\nu\right] >-1, \re\left[\mu\right] >-1$,  we
may perform the trivial $k$-integration (due to the
$\delta$-function) to finally put (\ref{Q_a,z_1}) in the form
\beq
\label{Qaz}
\Q_{a,z} = -{1\over 2}\sum_{b<a}\hbar c\vert k_{ab}\vert\Gamma_{a\rightarrow b}^{\,z}\Bigg\lbrace {2\over 3}-G_{\parallel}\left(2\vert k_{ab}\vert\rho\sin\phi\right) - \sum_{l=1}^{q-1}\left[G_{\parallel}\left(2\vert k_{ab}\vert\rho\sin\left(\phi+\pi l/q\right)\right)-G_{\parallel}\left(2\vert k_{ab}\vert\rho\sin\left(\pi l/q\right)\right)\right]\Bigg\rbrace \, , \;\;
\eeq
\end{widetext}
where
\beq
\label{Gparal}
G_{\parallel}(x) = {\sin x\over x}+{\cos x\over x^2}-{\sin x\over x^3}
\eeq
and we have made use of the explicit expressions for $J_{1/2}(x)$ and
$J_{3/2}(x)$. We have also defined
\begin{equation}
\label{EElivre}
\Gamma^{j}_{a\rightarrow b} = \frac{4}{\hbar} \vert \langle a \vert d_{j} \vert b \rangle \vert^2 \vert k_{ab} \vert^3 \, ,
\end{equation}
which is nothing but the spontaneous emission rate between levels a and b for
the $j$-polarization in free space, assuming of course that $a > b$.

The calculations for $\Q_{a,\rho}$ and $\Q_{a,\phi}$
are a little bit more involved but rather analogous to the one we
just showed. In these cases the m-summations that we need to
evaluate are
\beq
\label{sum_qm2J2}
\sum_{m=-\infty}^{\infty} m^2 J_{qm}^2\left(\kappa\rho\right) \bigg\{
\begin{array}{c}
\cos^2\left(qm\phi\right) \\
\sin^2\left(qm\phi\right)
\end{array} \bigg\} \, ,
\\
\label{sum_Jqm'2}
\sum_{m=-\infty}^{\infty}J_{qm}^{\,\prime\,2}\left(\kappa\rho\right) \bigg\{
\begin{array}{c}
\cos^2\left(qm\phi\right) \\
\sin^2\left(qm\phi\right)
\end{array} \bigg\} \, .
\eeq
\begin{widetext}
The first ones, shown in (\ref{sum_qm2J2}), may be carried out by considering the particular case
of (\ref{BesselGAT}) in which $Z_{\nu + qm}(\kappa \rho) =
J_{qm}(\kappa \rho)$ and differentiating it twice with respect to
the variable $\phi$. After some algebraic manipulations we get
\beq
\label{qm2J2qm}
\hspace{-8pt}\frac{q^2}{\kappa^2\rho^2} \sum_{m=-\infty}^{\infty} m^2 J_{qm}^2\left(\kappa\rho\right)\cos\left(2qm\phi\right) = {1\over q}\sum_{l=0}^{q-1}\left[J_0\left(2\kappa\rho\sin\psi_l\right)\cos^2\psi_l-{J_1\left(2\kappa\rho\sin\psi_l\right)\over2\kappa\rho\sin\psi_l}\right] \, ,
\eeq
where $\psi_l = \phi + \pi l/q$, which immediately leads to
\beq
\label{qm2J2qm_sin}
\sum_{m=-\infty}^{\infty} m^2 J_{qm}^2\left(\kappa\rho\right)
\sin^2\left(qm\phi\right)= \frac{1}{2q}
\sum_{l=0}^{q-1}\hspace{-8pt}&&\left[-J_0\left(2\kappa\rho\sin\psi_l\right)\cos^2\psi_l + J_0\left(2\kappa\rho\sin\pi l/q\right)\cos^2\pi l/q \right. \nonumber \\ \hspace{-8pt}&&\left. + \frac{J_1\left(2\kappa\rho\sin\psi_l\right)}{2\kappa\rho\sin\psi_l} - \frac{J_1\left(2\kappa\rho\sin\pi l/q\right)}{2\kappa\rho\sin\pi l/q} \right] \, ,
\\
\label{qm2J2qm_cos}
\sum_{m=-\infty}^{\infty} m^2 J_{qm}^2\left(\kappa\rho\right)
\cos^2\left(qm\phi\right)= \frac{1}{2q}
\sum_{l=0}^{q-1}\hspace{-8pt}&&\left[J_0\left(2\kappa\rho\sin\psi_l\right)\cos^2\psi_l + J_0\left(2\kappa\rho\sin\pi l/q\right)\cos^2\pi l/q \right. \nonumber \\ \hspace{-8pt}&&\left. - \frac{J_1\left(2\kappa\rho\sin\psi_l\right)}{2\kappa\rho\sin\psi_l} - \frac{J_1\left(2\kappa\rho\sin\pi l/q\right)}{2\kappa\rho\sin\pi l/q} \right] \, .
\eeq
The remaining summations in (\ref{sum_Jqm'2}) may be evaluated by making use of the identity
\beq
\label{identity}
J_{qm}^{\,\prime\,2}\left(\kappa\rho\right)
&=& \frac{qm}{\kappa \rho}
J_{qm}^{\,\prime}\left(\kappa\rho\right)J_{qm}\left(\kappa\rho\right)
- \partial_{\kappa \rho} \left[
J_{qm}\left(\kappa\rho\right)J_{qm+1}\left(\kappa\rho\right) \right] \nonumber \\
&+& J_{qm}^2\left(\kappa\rho\right)
- \frac{qm+1}{\kappa \rho}
J_{qm}\left(\kappa\rho\right)J_{qm+1}\left(\kappa\rho\right) \, ,
\eeq
plus the particular cases of (\ref{BesselGAT}) for $Z_{\nu + qm}(\kappa \rho) = J_{qm}(\kappa \rho) , J_{qm+1}(\kappa \rho)$ and its derivatives with respect to variables $\rho$ and $\phi$. This yields
\beq
\label{Jqm'2_sin}
\sum_{m=-\infty}^{\infty}J_{qm}^{\,\prime\,2}\left(\kappa\rho\right)\sin^2\left(qm\phi\right)={1\over 2q}\sum_{l=0}^{q-1}\hspace{-8pt}&&\left[J_0\left(2\kappa\rho\sin\psi_l\right)\sin^2\psi_l - J_0\left(2\kappa\rho\sin\pi l/q\right)\sin^2\pi l/q \right. \nonumber \\
\hspace{-8pt}&&\left. - {J_1\left(2\kappa\rho\sin\psi_l\right)\over2\kappa\rho\sin\psi_l} + {J_1\left(2\kappa\rho\sin\pi l/q\right)\over2\kappa\rho\sin\pi l/q} \right] \, ,
\\
\label{Jqm'2_cos}
\sum_{m=-\infty}^{\infty}J_{qm}^{\,\prime\,2}\left(\kappa\rho\right)\cos^2\left(qm\phi\right)={1\over 2q}\sum_{l=0}^{q-1}\hspace{-8pt}&&\left[-J_0\left(2\kappa\rho\sin\psi_l\right)\sin^2\psi_l - J_0\left(2\kappa\rho\sin\pi l/q\right)\sin^2\pi l/q \right. \nonumber \\
\hspace{-8pt}&&\left.+ {J_1\left(2\kappa\rho\sin\psi_l\right)\over2\kappa\rho\sin\psi_l} + {J_1\left(2\kappa\rho\sin\pi l/ql\right)\over2\kappa\rho\sin\pi l/q} \right] \, .
\eeq
By inserting equations (\ref{qm2J2qm_sin}) - (\ref{Jqm'2_cos}) into the expressions for $\Q_{a,\rho}, \Q_{a,\phi}$ and following the same procedure that led to (\ref{Qaz}), we finally get
\beq
\label{Qaphi}
\Q_{a,\phi}&=&-{1\over 2}\sum_{b<a}\hbar c\vert k_{ab}\vert\Gamma_{a\rightarrow b}^{\,\phi}\Bigg\lbrace {2\over 3}-H_{\phi}\left(2\vert k_{ab}\vert\rho,\phi\right) - \sum_{l=1}^{q-1}\left[H_{\phi}\left(2\vert k_{ab}\vert\rho,\phi+\pi l/q\right)+H_{\phi}\left(2\vert k_{ab}\vert\rho,\pi l/q\right)\right]\Bigg\rbrace \, , \\
\label{Qarho}
\Q_{a,\rho}&=&-{1\over 2}\sum_{b<a}\hbar c\vert k_{ab}\vert\Gamma_{a\rightarrow b}^{\,\rho}\Bigg\lbrace {2\over 3}-H_{\rho}\left(2\vert k_{ab}\vert\rho,\phi\right) - \sum_{l=1}^{q-1}\left[H_{\rho}\left(2\vert k_{ab}\vert\rho,\phi+\pi l/q\right)-H_{\rho}\left(2\vert k_{ab}\vert\rho,\pi l/q\right)\right]\Bigg\rbrace \, ,
\eeq
\end{widetext}
where
\beq
\label{H1}
\vspace{-0.2in} H_{\phi}\left(x,\psi\right)&=&G_{\parallel}\left(x\sin\psi\right)\sin^2\psi \nonumber \\
\vspace{-0.1in} && \hspace{0.3in} + 2G_{\perp}\left(x\sin\psi\right)\cos^2\psi \, \\
\label{H2}
\vspace{-0.2in} H_{\rho}\left(x,\psi\right)&=&G_{\parallel}\left(x\sin\psi\right)\cos^2\psi \nonumber \\
\vspace{-0.1in} && \hspace{0.3in} + 2G_{\perp}\left(x\sin\psi\right)\sin^2\psi \, ,
\eeq
and also
\beq
\label{Gper}
G_{\perp}(x) = {\cos x\over x^2}-{\sin x\over x^3} \, .
\eeq

As we now have well suited expressions for the decaying rates, we
may proceed with some numerical investigations. In order to simplify
our discussion, let us assume that our atom has only one dominant
transition, so that we can neglect all terms in the summation of
states present in (\ref{Qaz})-(\ref{Qarho}) except for this dominant
term.

\begin{widetext}

\begin{figure}[!h]
\begin{center}
\scalebox{0.45}{\includegraphics{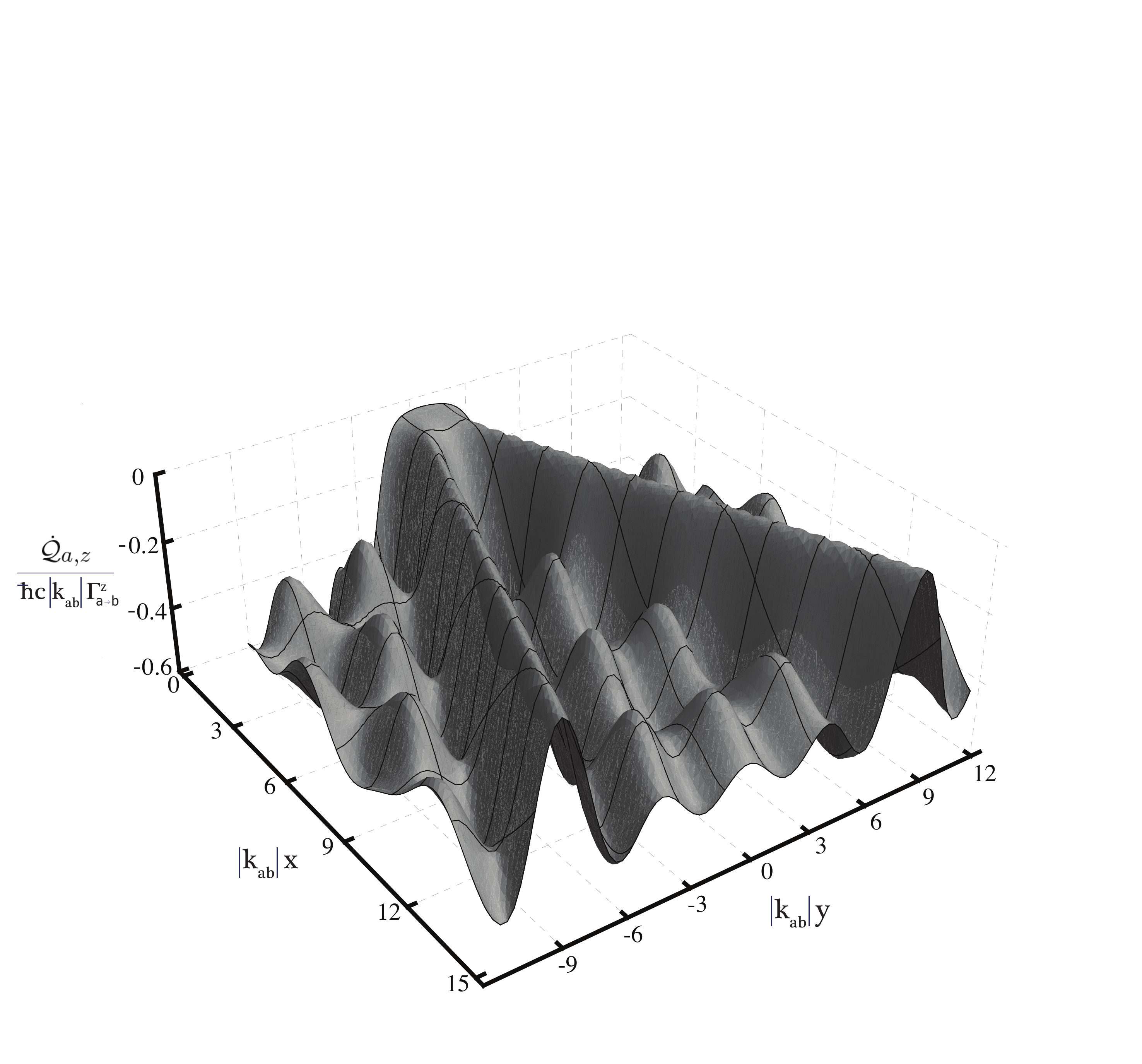}}
\end{center} \vspace{-0.3in}
\caption{The spontaneous emission for the $z$-polarization in a
wedge of $\pi/3$ radians. The axes {\small $\cal{X}, \cal{Y}$} and {\small$\cal{Z}$} follow the same
conventions of figure 1. The plates are
not displayed but the plot itself is very suggestive - they would be
sharply cutting the two highest slopes in half along the z
direction. Their aspect can be easily inferred from the diffuse
V-shape drawn by those slopes.}
\label{EEz3D}
\end{figure}



\begin{figure}[!h]
\begin{center}
\scalebox{0.45}{\includegraphics{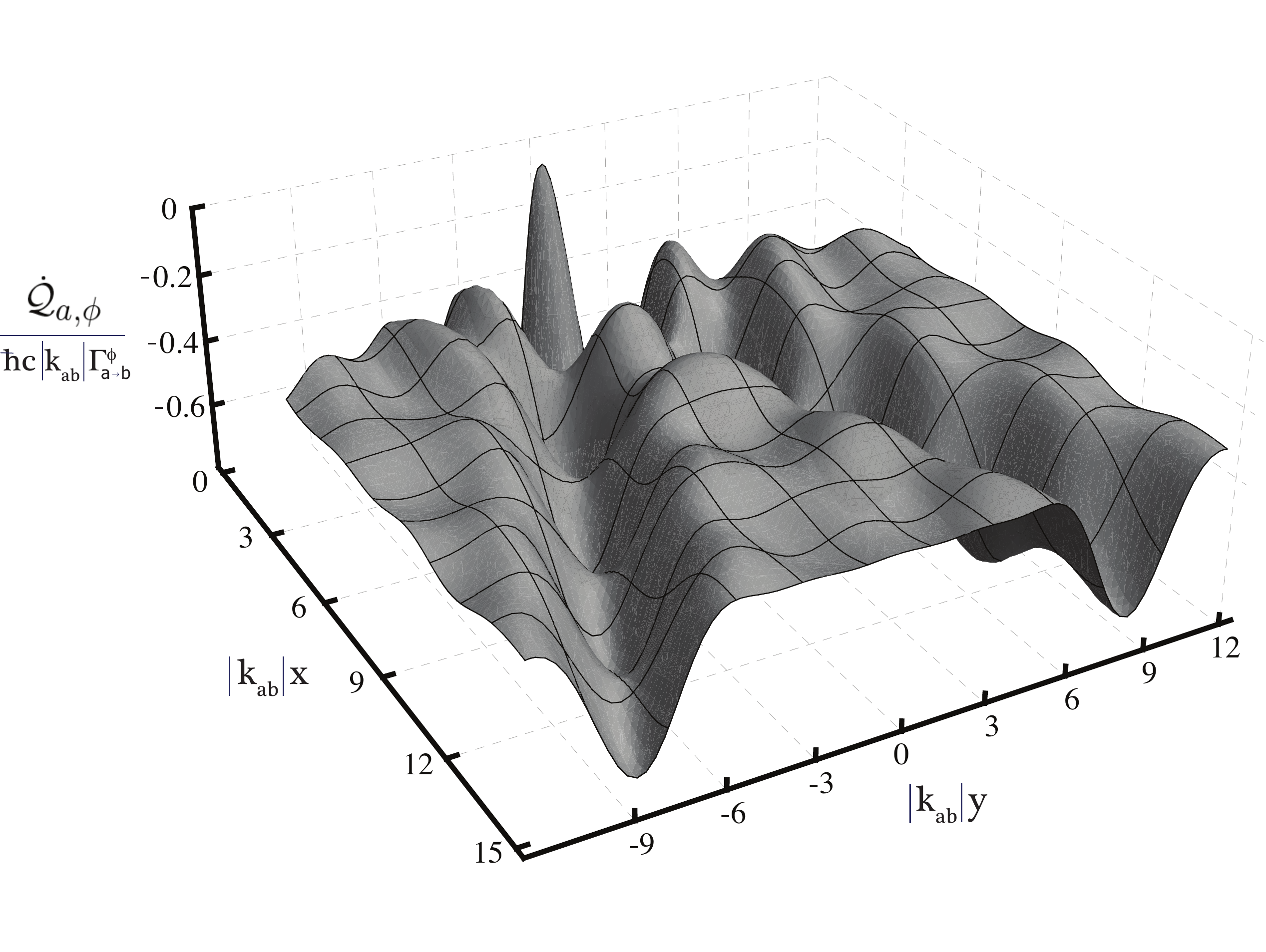}}
\end{center}  \vspace{-0.3in}
\caption{The spontaneous emission for the $\phi$ polarization
 in the same wedge as above.}
\label{EEphi3D}
\end{figure}

\end{widetext}

The first thing that we would like to point out is how
differently each polarization responds to the presence of the wedge.
By one hand, as one can see from Fig. \ref{EEz3D}, the
contribution of the $z$-polarization for the spontaneous emission
inside a wedge is greatly suppressed in the vicinity of the plates.
However, by taking a look at the equivalent graph for the
$\phi$-polarization shown in Fig. \ref{EEphi3D}, one concludes
that just the opposite occurs, which means that this contribution is
enhanced near the plates. This effect may seem curious initially,
but it is readily understood by recalling the boundary conditions
imposed by the wedge on the electric field, given by
${\bf\hat{\phi}} \times {\bf E}|_{S} = 0$. This implies that very
close to one of the plates and relatively away from the cusp the
electric field is approximately perpendicular to the mirror surface.
Since we are working in the dipole approximation, we conclude that
the parallel component of the dipole no longer couples to the field
while the perpendicular component coupling is maximized. This
explains both the vanishing of $\Q_{a,z}$ (and also of
$\Q_{a,\rho}$) and the enhancement of $\Q_{a,\phi}$ near the plates. We should mention that the previous reasoning is not valid when that atom is too close to the cusp, and we see accordingly that in such a case the SE vanishes for {\it both} polarizations. This is, of course, a direct consequence of the vanishing of the electric field at the cusp, as it may be seen from expressions (\ref{TM}) and (\ref{TE}).  

Intuitively, we should expect that a wedge
characterized by very small angles should mimic the behavior of two
parallel plates, provided the atom is not too close to the cusp. The
fact that we have checked numerically that our expressions coincide
with the known results for parallel plates in the limit $\alpha
\rightarrow 0$, $\rho \rightarrow \infty,\rho\alpha \rightarrow
const.$, encouraged us to consider less extreme situations, in which $\alpha$
is small but not infinitesimal. An interesting thing happens when
the setup under consideration is like the one depicted in Fig. 4, in
which an atom with $\Q_{a,\phi} = 0$ is placed in the $X$ axis at a
given distance from the cusp $\rho=x$. If $\alpha \ll 1$ it is
reasonable to approximate the wedge by two parallel plates, but the
distance between these effective plates changes as we move the atom
along the $X$ axis, like illustrated in Fig 5. So,
by considering an atom in different positions at the bisecting plane
of an acute wedge, we are approximately also looking to the
situation in which the atom is halfway between two parallel plates
and these plates are put at different separations.   In Fig. 6 we
show the behavior of $\Q_{a,\|} = \Q_{a,z}+\Q_{a,\rho}$ as we move the atom
along the $X$ axis for different values of $q$. We see that all
curves are practically zero up to approximately $k_{ab}x = q$, where
they rise abruptly and then start oscillating around a decaying mean
value. This is more clearly understood in terms of effective parallel
mirrors, as we shall see.


\begin{figure}[h!]
\begin{center}
\scalebox{0.45}{\includegraphics{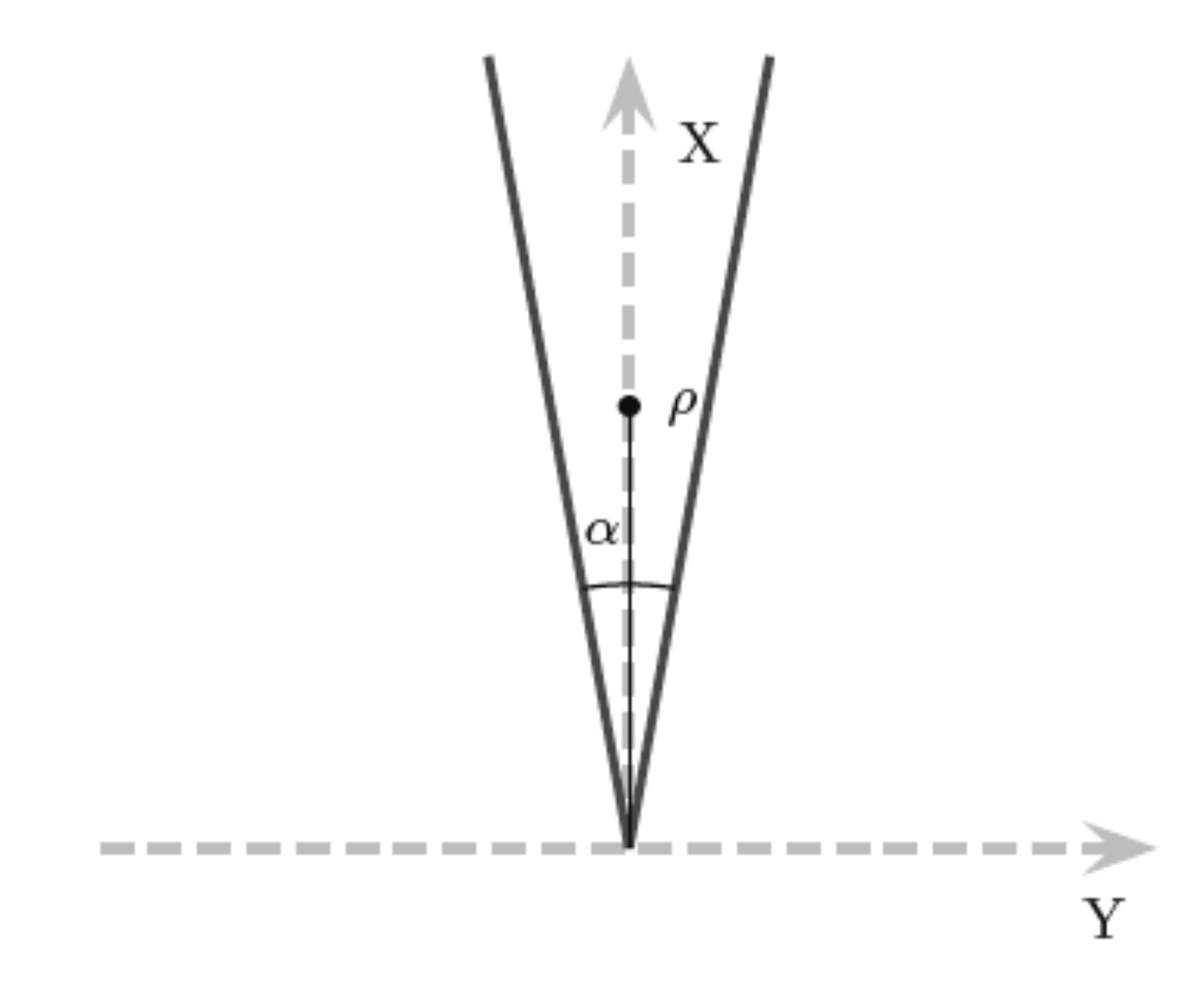}}
\caption{An atom localized in the bisecting plane of a wedge}
\end{center}
\label{Fig 4}
\end{figure}
%
%

The perpendicular component of the wave vector characterizing the modes between two plates is discrete and given by $n\pi/\delta$, where $\delta$ is the distance between the plates. When $k_{ab}x < q$, we have necessarily $\delta < \lambda_{ab}/2$
(reminding that $\delta \approx x\alpha = x \pi/q$ and $\lambda_{ab}
= 2 \pi / k_{ab}$). This implies that all the modes with $n>0$ are
more energetic then the $a \rightarrow b$ transition, making their
excitation by the dipole impossible. That said, we conclude that the
only way in which the dipole could emit would be to populate the
mode $n=0$, but that is also impossible since the zero mode is
polarized perpendicularly to the plates \cite{Hinds-LesHouches}, not
coupling to the dipole at all \cite{footnote3}. The situation changes dramatically once $k_{ab}x \approx q$, since the availability of at least one
mixed (with perpendicular and parallel components) mode with
$E_{mode} < \hbar \omega_{ab}$ allows the atom to emit, explaining the abrupt jump at that point. Based on the same reasoning, we should expect similar jumps each time the distance between the atom and the cusp reaches a multiple of $\lambda_{ab}/2$ (or equivalently, if $k_{ab} x$ reaches a multiple of $q$), but as we see from the curves corresponding to $q=60$ and $q=90$, there are jumps at $k_{ab}x \approx 3q$ but not at $k_{ab}x \approx 2q$. This may seem strange at first sight, but it is a direct consequence of the structure of the allowed modes between two mirrors. As we said before, these modes are characterized by an integer $n$ present in the perpendicular component of the respective wave vectors. By using the explicit form of these modes \cite{Hinds-LesHouches}, it is possible to show that the ones associated with even integers have a node at $\delta/2$, producing a vanishing electric field there. Since our atom is effectively located at $\delta/2$, we conclude that it feels only the influence of the odd modes, and that is the reason for the jumps only at $k_{ab} x$ equal to odd values of $q$.


\begin{figure}[h!]
\begin{center}
\scalebox{0.45}{\includegraphics{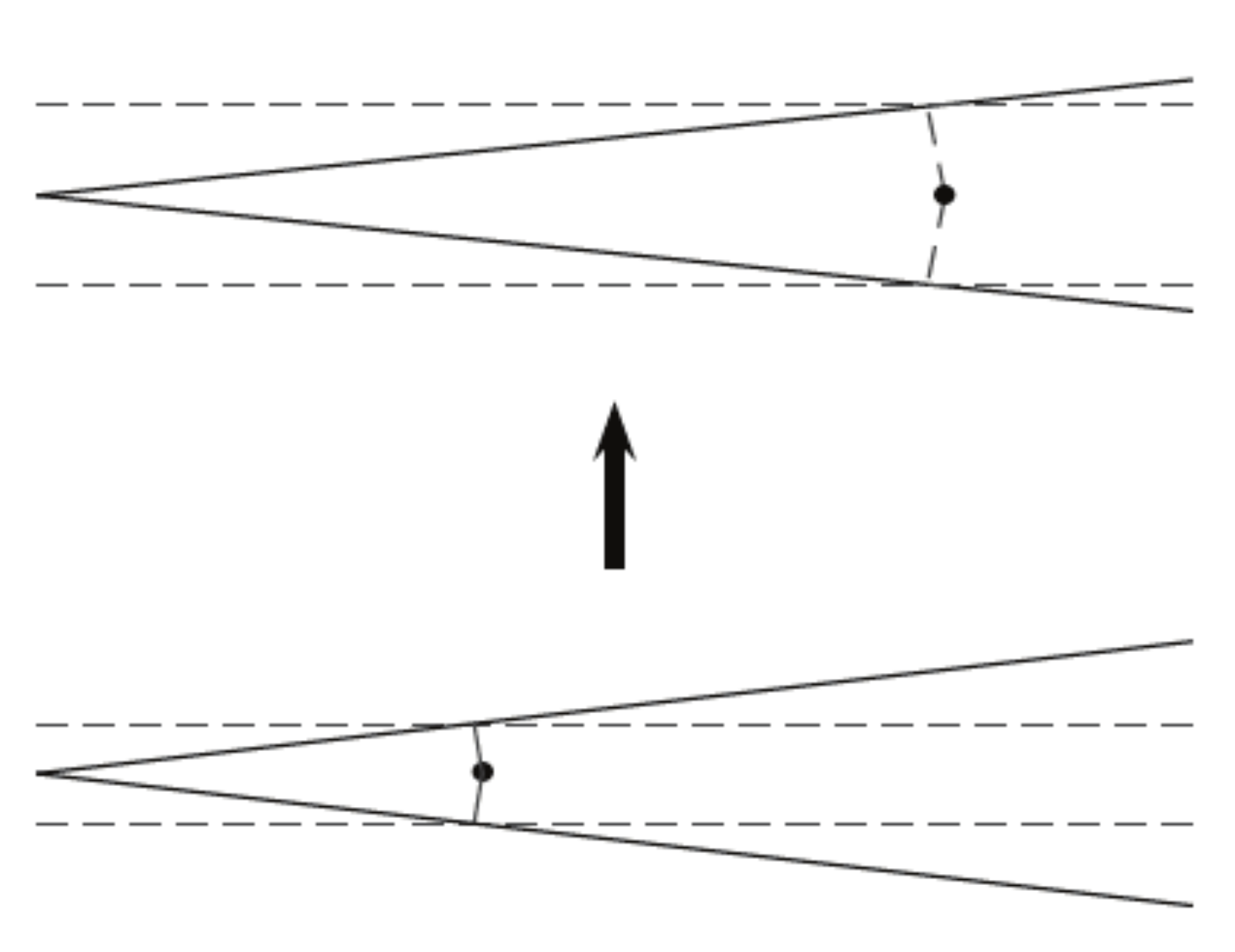}}
\caption{Effective mirrors at two different positions of the atom. We note that the distance between the effective plates becomes larger as the atom moves away from the cusp.}
\end{center}
\label{Fig 5}
\end{figure}

%
%

We shall close this work by considering a limiting case
of our expressions, in which we put $\alpha=\pi$ and the wedge
degenerates into a single plate. Then $q=1$ and by evaluating (\ref{Qaz}), (\ref{Qaphi}) and (\ref{Qarho}) at $\phi = \pi/2$ we get
\beq
\Q_{a,\|} = \Q_{a,x} + \Q_{a,z} = -\frac{1}{2} \hbar k_{ab} c (\Gamma^{x}_{a\rightarrow b} + \Gamma^{z}_{a\rightarrow b}) \cdot \nonumber \\
 \left[\frac{2}{3} - \frac{\sin(2\vert k_{ab}\vert x)}{2\vert k_{ab}\vert x} - \frac{\cos (2\vert k_{ab}\vert x)}{(2\vert k_{ab}\vert x)^2} + \frac{\sin (2\vert k_{ab}\vert x)}{(2\vert k_{ab}\vert  x)^3} \right] \, ,
\eeq
\beq
\Q_{a,y} = - \hbar k_{ab} c  \Gamma^{y}_{a\rightarrow b}  \left[ \frac{1}{3} - \frac{\cos (2\vert k_{ab}\vert x)}{(2\vert k_{ab}\vert x)^2}  \right. \nonumber \\ \vspace{-0.1in} \left. + \frac{\sin (2\vert k_{ab}\vert x)}{(2\vert k_{ab}\vert  x)^3} \right] \, ,
\eeq
which is in precise agreement with the known result from the literature \cite{Milonni, Hinds-LesHouches}.

\section{Conclusion and perspectives}

In this paper we investigated how the spontaneous emission rate of
an atom is affected by the presence of a conducting wedge. We used a
very general method, which allows for the evaluation of van der
Waals and excited potentials as well as spontaneous emission rates.
Firstly we applied it to the problem of an atom inside a mirror
wedge. In the following section we showed that, although for a wedge
of an arbitrary angle $\alpha$ the expressions are considerably
involved, when $q \in \mathbb{N}$ a lot of simplification was
possible and we arrived at manageable expressions for SE rates. We
then obtained the contribution of each polarization and plotted two
of them separately. Note that the behavior of the SE rate in a wedge
shows the usual oscillatory pattern.

As expected, the phenomenon of suppression of the SE
rate can also occur for an atom inside a wedge. As we showed
graphically, for very small angles of the wedge, the plates forming
it behave effectively as they were two parallel plates. Hence, for
polarizations parallel to these effective plates, namely,
polarizations $\rho$ and $z$, there are configurations for which the
atom will not decay at all.

There are several possible generalizations for this work, the most
obvious one being how to bring some real effects into consideration,
like finite conductivity and temperature corrections, which fall out
of the scope of this paper but serve as a guide for future research.

\begin{widetext}

\vspace{-0.3in}
\begin{figure}[!h]
\begin{center}
\scalebox{0.475}{\includegraphics{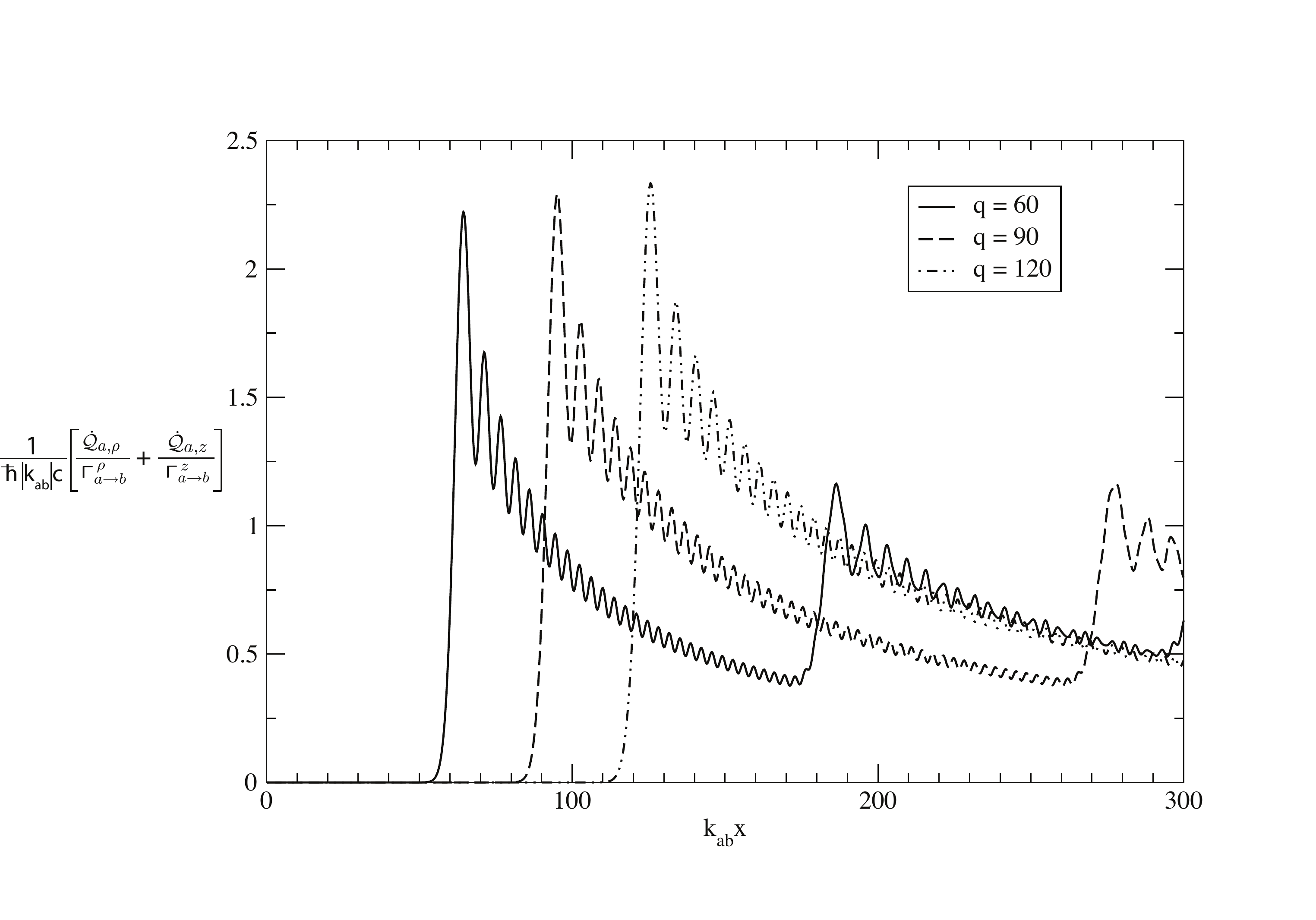}}
 \end{center}  
\vspace{-0.1in}
\caption{Spontaneous emission for an atom polarizable only
 in the $\rho$ and $z$ directions and located along the bisector
 plane of the wedge as a function of its distance from the cusp.
 Note the occurrence of the phenomenon of suppression for distances
 from the cusp smaller than a certain value determined by the angle of the wedge, followed by a sudden jump. In the curves corresponding to $q=60$ and $q=90$ it is even possible to see further jumps (at $k_{ab}x \approx 180$ for $q=60$ and $k_{ab}x \approx 270$ for $q=90$), caused by the sudden availability of extra modes (see also the text). }
 \label{EEphi2D}
 \end{figure}
 
 \end{widetext}

{\bf Acknowledgments}: FSSR and CF are very indebted to G. Barton,
R. Passante and R. Rodrigues for valuable criticism and comments.
FSSR would like to thank M. Moriconi for the kind hospitality at
Universidade Federal Fluminense, where part of of this work was
developed, and also acknowledge FAPERJ for parcial financial
support. Furthermore, TCNM acknowledges FAPERJ for financial support
and CF thanks CNPq for partial financial support.


\begin{thebibliography}{99}
\bibitem{footnote1} We should mention that these are not the
 only novelties in this work. In this same paper Einstein discusses many
  other things, like the properties of the photon, which also are of key
   importance. For a better account of this paper accomplishments,
   see \cite{Kleppner, Milonni}.
\bibitem{footnote2} Actually the role of boundaries in the SE was a disputed
 matter sometime ago, see section 6.2 of \cite{Milonni}.
\bibitem{footnote3} Incidentally, for parallel polarizations this picture also gives the right result near the cusp, where the approximation of parallel plates should no longer hold.
\end{thebibliography}
\end{document}